\documentstyle[12pt]{article}

\def\p{\partial}
\def\be{\begin{equation}}
\def\ee{\end{equation}}
\def\ben{\begin{displaymath}}
\def\een{\end{displaymath}}
\def\ba{\begin{array}{c}}
\def\ea{\end{array}}

\begin{document}

 \begin{center}

{\Large \bf Pseudo-Hermitian approach to
\\
energy-dependent Klein-Gordon models }

\vspace{.5cm}

{ Miloslav Znojil,
  Hynek B\'{\i}la
    and V\'{\i}t Jakubsk\'{y} }

  {Nuclear Physics Institute, 250 68 \v{R}e\v{z}, Czech Republic}
%

PACS {03.65.Bz  03.65.Ca}     

\end{center}

\subsection*{Abstract}

The relativistic Klein-Gordon system is studied as an illustration
of Quantum Mechanics using non-Hermitian operators as observables.
A version of the model is considered containing a generic
coordinate- and energy-dependent phenomenological mass-term
$m^2(E,x)$. We show how similar systems may be assigned a pair of
the {\em linear},  energy-independent left- and right-acting
Hamiltonians with quasi-Hermiticity property and, hence, with the
standard probabilistic interpretation.

\bigskip

KEYWORDS: Quantum Mechanics;  energy-dependent forces; pseudo- and
quasi-Hermitian formalism; relativistic kinematics; linear
representation of observables


\bigskip

\section{Introduction}

In units $\hbar = c = 1$ Quantum Mechanics describes the motion of
a free spinless particle by the partial (parabolic) differential
Schr\"{o}dinger equation
 \be
 i\,\p_t \psi = -\frac{1}{2m}\,\triangle \psi
 \label{SE}
 \ee
or, in the relativistic kinematical regime, by the hyperbolic
Klein-Gordon equation
 \be
 \left (i\,\p_t \right )^2 \psi =
  \left (-\triangle + m^2 \right ) \psi\,.
  \label{KG}
 \ee
Various phenomenological requirements may be met via an
introduction of a suitable time-independent interaction which
still admits the usual formal Fourier-transformation separation of
the time-dependence.

We intend to contemplate a generic situation where the form of the
interaction is allowed to vary with the energy. A schematic
clarification of a few basic features of the models of this type
may be offered by the Schr\"{o}dinger equation (\ref{SE}) where we
may put $\psi=\psi(t)= e^{-iEt}\Psi(r)$ and postulate the most
elementary harmonic-oscillator form of the interaction. In one
dimension this leads to the ordinary differential equation
 \begin{equation}
 -\,\frac{1}{2m(E)}\ \frac{\rm d^2}{\rm dr^2}\, \Psi({r}) + {r}^2
\Psi({r})= E \Psi({r}) \label{hoSE}\
 \end{equation}
where a few possible effects of the variability of the mass with
the energy may be illustrated by the most elementary ansatz $ 2
m(E) = A^2 \,(E-E_0)^2$. A mere rescaling of eq. (\ref{hoSE})
leads to the solvable bound-state problem with the spectrum
determined by the closed formulae
 \begin{equation}
 E=\left\{
 \begin{array}{c}
 E_n^{(+)}=E_0+\sqrt{E_0^2+({8n+4})/{A}}, \ \ \ \ n = 0, 1,
 \ldots,\\
 E_{\pm n}^{(-)}=E_0\pm \sqrt{E_0^2-({8n+4})/{A}}, \ \ \ \ n = 0, 1,
 \ldots,n_{max}
 \end{array}
 \right . \,.
 \end{equation}
The unusual second family is finite and remains non-empty only for
$A\,E_0^2\geq 4$. New levels emerge at each new $n_{max}
=entier[(A\,E_0^2- 4)/8]$.

\subsection{Relativistic kinematics as a challenge in physics
}

Our intuition may fail in similar situations, indeed. In a way
complementing the recent study of nonrelativistic energy-dependent
descendants of eq. (\ref{SE}) \cite{lin}, we intend to pay
attention to the relativistic models where the standard
Klein-Gordon Hamiltonians (= the operators of energy) {\em do not
coincide} with the generators of time evolution \cite{FV}.

This is a challenging difficulty. In addition we see an even
stronger reason for interest in Klein-Gordon system in its close
connection to an extended Quantum Mechanics using non-Hermitian
operators  \cite{exper}. Thus, we are going to pay attention to
the relativistic Klein-Gordon equations
 \begin{equation}
 \
 \left( i\,\partial_t\right )^2 \Psi^{(KG)}(x,t)
 = \hat{H}^{(KG)}\,\Psi^{(KG)}(x,t)\,
 \label{jednar}
 \end{equation}
with interactions mediated by a coordinate- and energy-dependence
of the phenomenological mass-term in $\hat{H}^{(KG)} = -\triangle
+m^2(E,x)$.

\subsection{Non-Hermiticity of Hamiltonians as a challenge in
mathematics \label{tro}}

It is well known \cite{Most} that the usual Klein-Gordon
Hamiltonians $\hat{H}^{(KG)}$ are pseudo-Hermitian or, in a more
compact wording, $PT-$symmetric \cite{BB}. The latter type of
symmetry emerged in Quantum Mechanics as related to the imaginary
cubic (cf. \cite{Caliceti}) or negative quartic (cf. \cite{BG})
anharmonicities, with important implications expected also within
the relativistic Quantum Field Theory \cite{Bessis}. Nevertheless,
its key importance has only been revealed and emphasized by Bender
and Boettcher \cite{BB} who conjectured that the enigmatic reality
of spectra of certain nonrelativistic Hermiticity-violating
oscillators might be attributed to an unusual commutativity of
their Hamiltonians with the product $PT$ of parity ${\cal P}$ and
time reversal ${\cal T}$.

Later on, it became clear that all the similar models (with both
the purely real or mixed, complex-conjugate spectra) seem to obey
an unusual, pseudo-unitary time-evolution law \cite{cons}
reflecting the pseudo-Hermiticity of the Hamiltonian. One had to
return to the older work by Dirac et al \cite{Dirac} to imagine
that the self-adjoint and invertible operator ${\cal P}$ replaces
the usual identity operator in the role of an indeterminate
pseudo-metric in the Hilbert space. In some exactly solvable
examples the puzzling indeterminacy of the corresponding
pseudo-norm has been attributed to the mere sign-type variable
called quasi-parity \cite{ptho} or, later but much better, charge
$C$ \cite{BBJ}.

Ali Mostafazadeh \cite{AMos} realized that formally, there are no
real reasons against the replacement of the indeterminate
pseudo-metric ${\cal P}$ (or $\eta$ in his - or rather Dirac's -
preferred notation) by an ``equivalent" (though still in general
non-unique) positive-definite ``physical" metric $\eta_+$ which,
in essence, coincides with the operator ${\cal CP}$ of ref.
\cite{BBJ} and admits the current probabilistic interpretation of
the theory.

It is amusing to notice that at the moment of its introduction,
the operator $\eta_+>0$ itself has already been used and studied
for more than ten years, within the framework of nuclear physics.
In the review paper by Scholtz et al \cite{Geyer}, the name
``quasi-Hermitian" has been coined for all the ``physically
consistent" (i.e., in the present language, ${\cal CPT}-$symmetric
\cite{BBJ}) observables, i.e., for all the operators $H$ such that
$H^\dagger = \eta_+\,H\,\eta_+^{-1}$ in the notation of ref.
\cite{AMos}. As long as $\eta_+$ varies, in general, with the
Hamiltonian, its construction, trivial as it may seem in some
exactly solvable examples \cite{Quesne}, represents a really
formidable task in the majority of the complicated models of
Quantum Mechanics \cite{CMBhere} and Field Theory \cite{Jones}.

\section{Klein-Gordon models}

\subsection{Pseudo-Hermiticity}

We intend to show how the nonlinearity caused by the energy
dependence may be suppressed by the separation of the left- and
right-action of the Hamiltonian. In (\ref{jednar}) we abbreviate
$\Psi^{(KG)}(x,t)=\varphi_2^{(SR)}(x,t)$ and
$i\,\partial_t\,\Psi^{(KG)}(x,t)=\varphi_1^{(SR)}(x,t)$ and get
the following Schr\"{o}dinger-type re-arrangement of our equation,
 \ben
 i\,
 \partial_t
 \left (
 \begin{array}{c}
 \varphi_1^{(SR)}(x,t)\\ \varphi_2^{(SR)}(x,t)
 \end{array}
 \right )
=
 \hat{h}^{(SR)}\,
 \cdot \left (
 \begin{array}{c}
 \varphi_1^{(SR)}(x,t)\\ \varphi_2^{(SR)}(x,t)
 \end{array}
 \right ), \ \ \ \ \ \ \ \
 \hat{h}^{(SR)}=\left (
 \begin{array}{cc}
 0&\hat{H}^{(KG)}\\
 1&0
 \end{array}
 \right )
 \,.
 \label{jdvar}
 \een
As long as the concept of pseudo-Hermiticity of an operator $A$
requires just the fulfillment of the operator relation $A^\dagger
= \eta\,A\,\eta^{-1}$, the ``pseudo-metric" operator
$\eta=\eta^\dagger$ is, in general, indeterminate and
$A-$dependent. Still, we may immediately assume the
pseudo-Hermiticity of the energy operator,
 \ben
 \left (
 \hat{H}^{(KG)}
 \right )^\dagger =
 \eta^{(KG)}
 \hat{H}^{(KG)}
 \left (
 \eta^{(KG)}
 \right )^{-1}\,,
 \een
and note that it implies that
 \ben
 \left (
 \hat{h}^{(SR)}
 \right )^\dagger =
 \eta^{(SR)}
 \hat{h}^{(SR)}
 \left (
 \eta^{(SR)}
 \right )^{-1}, \ \ \ \ \ \ \ \ \
 \eta^{(SR)}=\left (
 \begin{array}{cc}
 0&\eta^{(KG)}\\
 \eta^{(KG)}&0
 \end{array}
 \right )
 \een
i.e., it already guarantees the pseudo-Hermiticity of the
generator $\hat{h}^{(SR)}$ of the time evolution.

\subsection{Quasi-Hermiticity}

In the spirit of paragraph \ref{tro} we must perform the second
step and replace both the generalized parities $\eta^{(SR)}\equiv
{\cal P}^{(SR)}$ and $\eta^{(KG)}\equiv {\cal P}^{(KG)}$ by the
respective positive, ``physical" metric operators $\eta^{(SR)}_+
\equiv {\cal C}^{(SR)}{\cal P}^{(SR)}$ and $\eta^{(KG)}_+ \equiv
{\cal C}^{(KG)}{\cal P}^{(KG)}$. This means that the corresponding
scalar products as well as the norms will behave in such a manner
that the axioms of Quantum Mechanics remain satisfied. In the
other words, our operators become ``Hermitian" whenever we decide
to understand the ``Hermiticity" in the new metric $\eta^{(SR)}_+
\equiv {\cal C}^{(SR)}{\cal P}^{(SR)}$. In this sense, also the
time-evolution of the Klein-Gordon system remains ``unitary" in
the new language.

\section{Separation of the left and right action of the
 Hamiltonians }

\subsection{Bi-orthogonal bases and energy as a parameter}

In any energy-dependent interaction term, we may tentatively
replace the variable energy $E$ by a fixed parameter $z$ and
compute the whole spectrum $E(z)$ of each $H(z)$ at any value of
$z$. This gives us a family of auxiliary spectral decompositions
 \be
 H_{}(z)= \sum_n\,|\Psi^{(n)}_{}(z)\rangle \,E^{(n)}_{}(z)
 \,\langle\Psi_{(n)}(z)|\,.
 \label{SE2}
 \ee
Thus, our pseudo-Hermitian input Hamiltonians $H(z)$ are, in the
light of the current textbooks \cite{Iochvidov}, easily tractable
as matrices in a suitable bi-orthogonal (or rather bi-orthonormal)
basis. This means that
 \ben
 \langle \Psi_{(m)} |\Psi^{(n)}\rangle = \delta_{m}^{n},
 \ \ \ \ \ \ \ m, n = 1, 2, \ldots \,
 \een
while the completeness relations may also be assumed in the form
of the infinite series,
 \be
\hat{I} = \sum_{n}\, |\Psi^{(n)}\rangle
   \langle \Psi_{(n)}|\,.
 \label{CoR}
 \ee
In the light of the pseudo-Hermiticity rules we may write
 \be
 H^\dagger \eta = \eta\,H=
  \sum_{n}\, |\Psi_{(n)} \rangle
  \,E^{(n)}\,
 \langle \Psi_{(n)}|, \ \ \ \ \ \ \ \ \ \
 \eta=
  \sum_{n}\, |\Psi_{(n)} \rangle
  \,\langle
 \Psi_{(n)}| = \eta^\dagger\,,
 \label{referli}
 \ee
 \be
 H \eta^{-1} = \eta^{-1}\,H^\dagger=
  \sum_{n}\, |\Psi^{(n)} \rangle
  \,E^{(n)}\,
 \langle \Psi^{(n)}|, \ \ \ \ \ \ \ \ \ \
 \eta^{-1}=
  \sum_{n}\, |\Psi^{(n)} \rangle
  \,\langle
 \Psi^{(n)}| \,.
 \label{referpo}
 \ee
Whenever the metric is positive, $\eta> 0$ we may call our
Hamiltonians  quasi-Hermitian. In an error-checking convention of
ref. \cite{lin} we may also temporarily consider only the formulae
where all the kets and bras are upper- and lower-indexed,
respectively.

\subsection{Innovated bi-orthogonal bases  \label{poHilbert}}

Once we return to the implicit constraint
 \be
 z_{phys} =E^{(n)}_{}(z_{phys})\,
 \label{nelin}
 \ee
and to its explicit solutions
 \ben
 z_{phys} =E^{(n,1)},E^{(n,2)},\ldots,   E^{(n,m(n))}\,
 \een
we have to move to a new basis. Thus, we denote
$|\Psi^{(n)}_{}(z_{phys})\rangle = |\phi^\alpha \rangle$ and
$|\Psi_{(n)}(z_{phys})\rangle = |\phi_\alpha \rangle$ and
abbreviate $E^{(n,j)}=E_\alpha$ using multi-indices
$\alpha=(n,j)$. All the overlaps are assumed calculable,
 \ben
 \langle \varphi_\alpha |\varphi^\beta \rangle =R_{\alpha}^{\ \beta}
 \,.
 \label{overlapspt}
 \een
This suffices for a formal representation of the unit projector,
 \be
\hat{I} = \sum_{\alpha, \beta\in A}\, |\varphi^\alpha \rangle
\left ( R^{-1} \right )_{\alpha}^{\ \beta}
 \langle \varphi_\beta|
 =
 \sum_{\beta\in A}\, |\varphi^\beta \rangle
 \rangle \,\langle \varphi_\beta|
 =
 \sum_{\alpha\in A}\, |\varphi^\alpha \rangle\,
 \langle \langle \varphi_\alpha|
 \,
 \label{CRpo}
 \ee
with abbreviations
 \ben
 \sum_{\alpha\in A}\, |\varphi^\alpha \rangle \left (
 R^{-1} \right )_{\alpha}^{\ \beta}
 \, \equiv\,
 |\varphi^\beta\rangle \rangle  \,,
 \ \ \ \ \ \ \ \ \ \ \ \ \
 \sum_{ \beta\in A}\,
 \left ( R^{-1} \right )_{\alpha}^{\ \beta}
 \langle \varphi_\beta| \, \equiv\,
 \langle \langle \varphi_\alpha| \,,
 \label{ufept}
 \een
 \ben
\langle \langle \varphi_\alpha |\varphi^\beta \rangle = \langle
\varphi_\alpha |\varphi^\beta \rangle\rangle =\delta_{\alpha}^{\
\beta}, \ \ \ \ \ \ \alpha, \beta \in A\,
 \een
and with the innovated completion relations
 \ben
\hat{I} = \sum_{\alpha\in A}\, |\varphi^\alpha \rangle \langle
 \langle \varphi_\alpha|=
  \sum_{\alpha\in A}\, |\varphi^\alpha \rangle \rangle
 \langle \varphi_\alpha|\,.
  \een

\section{The elimination of the energy dependence}

Two alternative tentative spectral expansions of our
quasi-Hermitian $H(E)$ read
 \ben
K = \sum_{\alpha \in A}\, |\varphi^\alpha \rangle
\,E_\alpha\,\langle
 \langle \varphi_\alpha|
\,,  \ \ \ \ \ \
 L = \sum_{\beta \in A}\, |\varphi^\beta \rangle
 \rangle\,E_\beta\,
 \langle \varphi_\beta|
 \,.
 \een
These operators share the action of $H(E)$ to the right and to the
left, respectively,
 \ben
 K\,|\varphi^\beta\rangle =E_\beta\,|\varphi^\beta\rangle\,,\ \ \ \
 \ \ \ \ \ \
 \langle \varphi_\alpha|\,L=E_\alpha\,\langle \varphi_\alpha|\,.
 \een
They are both energy-independent -- this is their main merit.
Their quasi-Hermiticity rules acquire the form
 \ben
 K^\dagger \mu = \mu\,K=
  \sum_{\alpha \in A}\, |\varphi_\alpha \rangle  \rangle
  \,E_\alpha\,\langle
 \langle \varphi_\alpha|,
 \ \ \ \ \ \ \ \ \ \
 \nu\, L = L^\dagger\,\nu=
  \sum_{\alpha \in A}\, |\varphi_\alpha \rangle
  \,E_\alpha\,
 \langle \varphi_\alpha|,
 \een
where one has to abandon the above-mentioned ``error-correcting"
convention,
 \ben
 \mu=
  \sum_{\alpha \in A}\, |\varphi_\alpha \rangle  \rangle
  \,\langle
 \langle \varphi_\alpha| = \mu^\dagger\,,
\ \ \ \ \ \ \ \ \ \
 \nu =
  \sum_{\alpha \in A}\, |\varphi_\alpha \rangle
  \,
 \langle \varphi_\alpha| = \nu^\dagger\,.
 \label{referwil}
 \een
 \ben
 \mu^{-1}=
  \sum_{\alpha \in A}\, |\varphi^\alpha \rangle
  \,
 \langle \varphi^\alpha| \,, \ \ \ \ \ \ \ \ \ \
 \nu^{-1} =
  \sum_{\alpha \in A}\, |\varphi^\alpha \rangle \rangle
  \,
 \langle \langle \varphi^\alpha| \,.
 \label{referwsa}
 \een
These Klein-Gordon related formulae complement and extend their
nonrelativistic predecessors of ref. \cite{lin}. We may summarize
our considerations by saying that the transition to the
relativistic kinematics requires a weaker (viz., pseudo- or
quasi-) Hermiticity of the initial (= Feshbach-Villars-type)
Hamiltonians $H^{(SR)}(E)$. The bra and ket vectors in the
spectral expansions {\em cease to be} the mere Hermitian
conjugates of each other because they are formed by the two sets
of eigenstates of $H$ and $H^\dagger$ \cite{lin}. Still, the work
in these bi-orthognal bases remains extremely natural, both before
and after the introduction of an energy dependence in our
interactions. In this sense, the {\em formal} differences between
the nonrelativistic and relativistic models appear to be minimal.

\section*{Acknowledgements}

Work supported by the AS CR project AV 0Z 104 8901 and by the GA
AS grant Nr. A 104 8302.

\end{document}